\documentclass{article}

\pdfoutput=1

\usepackage{amsmath}
\usepackage{amscd}
\usepackage{amsfonts}
\usepackage{float}
\usepackage{subfig, placeins}

\usepackage[longnamesfirst,sort]{natbib}

\usepackage{Sweave}
\begin{document}

\title{Sweave Documentation for ``Implementing Markov chain Monte Carlo: Estimating with confidence''}
\author{James M. Flegal and Galin L. Jones}
\maketitle

\section{Introduction}
This file is the Sweave documentation for the examples provided in \citet{fleg:jone:2010b}.  

\subsection{Batch Means}
The following function is required to calculate the variance estimators using batch means (BM) and overlapping batch means (OLBM).  We are thankful to Murali Haran whom wrote the original function to implement BM which we have expanded here.

\begin{Schunk}
\begin{Sinput}
> id <- function(x) return(x)
> mcse <- function(vals, bs = "sqroot", g = id, meth = "BM", warn = FALSE) {
+     N <- length(vals)
+     if (N < 1000) {
+         if (warn) 
+             cat("WARNING: too few samples (less than 1000)\n")
+         if (N < 10) 
+             return(NA)
+     }
+     if (bs == "sqroot") {
+         b <- floor(sqrt(N))
+         a <- floor(N/b)
+     }
+     else if (bs == "cuberoot") {
+         b <- floor(N^(1/3))
+         a <- floor(N/b)
+     }
+     else {
+         stopifnot(is.numeric(bs))
+         b <- floor(bs)
+         if (b > 1) 
+             a <- floor(N/b)
+         else stop("batch size invalid (bs=", bs, ")")
+     }
+     if (meth == "BM") {
+         Ys <- sapply(1:a, function(k) return(mean(g(vals[((k - 
+             1) * b + 1):(k * b)]))))
+         muhat <- mean(Ys)
+         sigmahatsq <- b * sum((Ys - muhat)^2)/(a - 1)
+         bmse <- sqrt(sigmahatsq/N)
+         return(bmse)
+     }
+     if (meth == "OBM") {
+         a <- N - b + 1
+         Ys <- sapply(1:a, function(k) return(mean(g(vals[k:(k + 
+             b - 1)]))))
+         muhat <- mean(Ys)
+         sigmahatsq <- N * b * sum((Ys - muhat)^2)/(a - 1)/a
+         bmse <- sqrt(sigmahatsq/N)
+         return(bmse)
+     }
+     else {
+         stop("method specified invalid (meth=", meth, ")")
+     }
+ }
\end{Sinput}
\end{Schunk}

\section{Normal AR(1) Markov Chains}
Consider the normal AR(1) time series defined by
\begin{equation} \label{eq:ar1}
   X_{n + 1} = \rho X_{n} + \epsilon_{n}
\end{equation}
where the $\epsilon_{n}$ are i.i.d.\ N(0,1) and $\rho < 1$.  This Markov chain has invariant distribution $\text{N}\left( 0 , 1/ (1 - \rho^2) \right)$.   

In our example, we perfomred calulcations for $\rho \in \{ 0.5, 0.95 \} $.

\subsection{Markov Chain Sampler}

The following chunk of code gives general functions needed to sample from \eqref{eq:ar1}.  Here we have a function that provides an observation from the chain, an observation q steps ahead with a defualt of one step ahead, and p observations from the chain.

\begin{Schunk}
\begin{Sinput}
> ar1 <- function(m, rho, tau) {
+     rho * m + rnorm(1, 0, tau)
+ }
> ar1.q <- function(m, rho, tau, q = 1) {
+     for (i in 1:q) {
+         m <- rho * m + rnorm(1, 0, tau)
+     }
+     m
+ }
> ar1.gen <- function(mc, p, rho, tau, q = 1) {
+     loc <- length(mc)
+     junk <- double(p)
+     mc <- append(mc, junk)
+     for (i in 1:p) {
+         j <- i + loc - 1
+         mc[(j + 1)] <- ar1(mc[j], rho, tau)
+     }
+     return(mc)
+ }
\end{Sinput}
\end{Schunk}

\subsection{Additional Functions}

The following are additional functions necessary for later calculations.  The first calculates the estimated first and third quartile while the second calculates the associated MCSE via subsampling.  Notice this function is similar to the function necessary for OLBM.

\begin{Schunk}
\begin{Sinput}
> quant <- function(input) {
+     quantile(input, prob = c(0.25, 0.75), type = 1)
+ }
> subsampling <- function(vals) {
+     N <- length(vals)
+     b <- floor(sqrt(N))
+     a <- N - b + 1
+     Ys <- sapply(1:a, function(k) return(quant(vals[k:(k + b - 
+         1)])))
+     muhat <- apply(Ys, 1, mean)
+     sigmahatsq <- N * b * apply((Ys - muhat)^2, 1, sum)/(a - 
+         1)/a
+     bmse <- sqrt(sigmahatsq/N)
+     return(bmse)
+ }
\end{Sinput}
\end{Schunk}

\subsection{Simulation Settings and Calculations}
In this next chunk of code, we first give the simulation settings used throughout the paper.  We then generate the two Markov chains for $\rho \in \{ 0.5, 0.95 \} $ and calculate the corresponding estimates and MCSEs.

\begin{Schunk}
\begin{Sinput}
> n <- 2000
> iter <- seq(1, n)
> crit.bm <- qt(0.9, (sqrt(iter) - 1))
> crit.obm <- qt(0.9, (iter - sqrt(iter)))
> set.seed(1976)
> rho1 <- 0.5
> rho2 <- 0.95
> chain1 <- ar1.gen(1, (n - 1), rho1, 1)
> mean1 <- cumsum(chain1)/seq(along = chain1)
> bm.est1 <- sapply(1:n, function(k) return(mcse(chain1[1:k], meth = "BM")))
> obm.est1 <- sapply(1:n, function(k) return(mcse(chain1[1:k], 
+     meth = "OBM")))
> quartile1 <- sapply(1:n, function(k) return(quant(chain1[1:k])))
> q.mcse1 <- sapply(1:n, function(k) return(subsampling(chain1[1:k])))
> chain2 <- ar1.gen(1, (n - 1), rho2, 1)
> mean2 <- cumsum(chain2)/seq(along = chain2)
> bm.est2 <- sapply(1:n, function(k) return(mcse(chain2[1:k], meth = "BM")))
> obm.est2 <- sapply(1:n, function(k) return(mcse(chain2[1:k], 
+     meth = "OBM")))
> quartile2 <- sapply(1:n, function(k) return(quant(chain2[1:k])))
> q.mcse2 <- sapply(1:n, function(k) return(subsampling(chain2[1:k])))
\end{Sinput}
\end{Schunk}

\subsection{Initial Examination of Output}
The following chunk of code will create plots for the initial examination of output.  The plots are also repeated in the document.

\begin{Schunk}
\begin{Sinput}
> rho = rho1
> chain <- chain1
> mean <- mean1
> par(mfrow = c(3, 1), mar = c(3, 4, 4, 2))
> ts.plot(chain, main = "Time-Series vs. Iteration", xlab = "", 
+     ylab = "", xlim = c(0, n))
> abline(h = 2 * sqrt(1/(1 - rho^2)))
> abline(h = -2 * sqrt(1/(1 - rho^2)))
> acf(chain, main = "Autocorrelation vs. Lag", ylab = "", xlab = "")
> ts.plot(mean, main = "Running Average vs. Iteration", xlab = "", 
+     ylab = "", lwd = 2, xlim = c(0, n))
> abline(h = 0)
> par(mfrow = c(1, 1), mar = c(5, 4, 4, 2))
\end{Sinput}
\end{Schunk}
\includegraphics{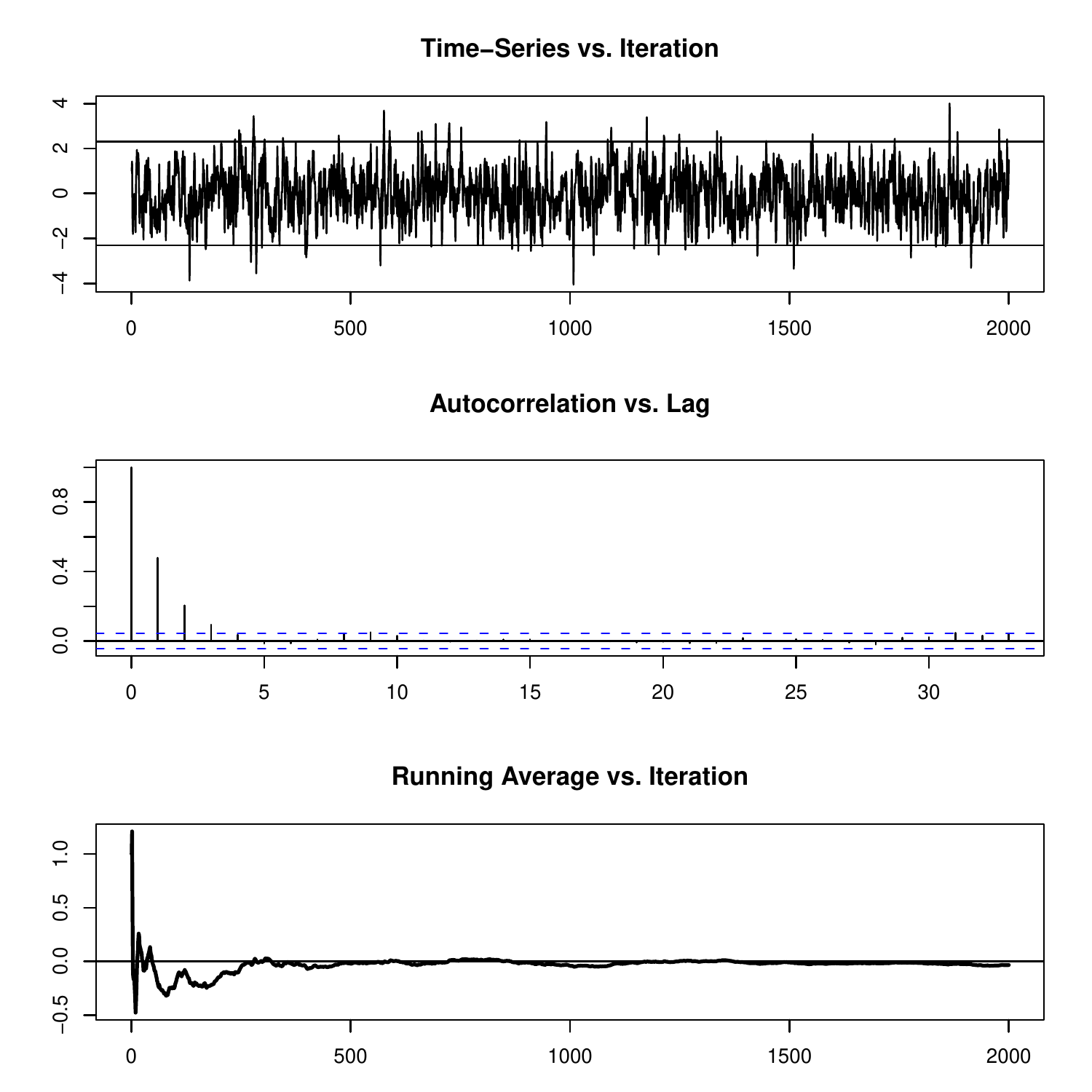}

\begin{Schunk}
\begin{Sinput}
> rho = rho2
> chain <- chain2
> mean <- mean2
> par(mfrow = c(3, 1), mar = c(3, 4, 4, 2))
> ts.plot(chain, main = "Time-Series vs. Iteration", xlab = "", 
+     ylab = "", xlim = c(0, n))
> abline(h = 2 * sqrt(1/(1 - rho^2)))
> abline(h = -2 * sqrt(1/(1 - rho^2)))
> acf(chain, main = "Autocorrelation vs. Lag", ylab = "", xlab = "")
> ts.plot(mean, main = "Running Average vs. Iteration", xlab = "", 
+     ylab = "", lwd = 2, xlim = c(0, n))
> abline(h = 0)
> par(mfrow = c(1, 1), mar = c(5, 4, 4, 2))
\end{Sinput}
\end{Schunk}
\includegraphics{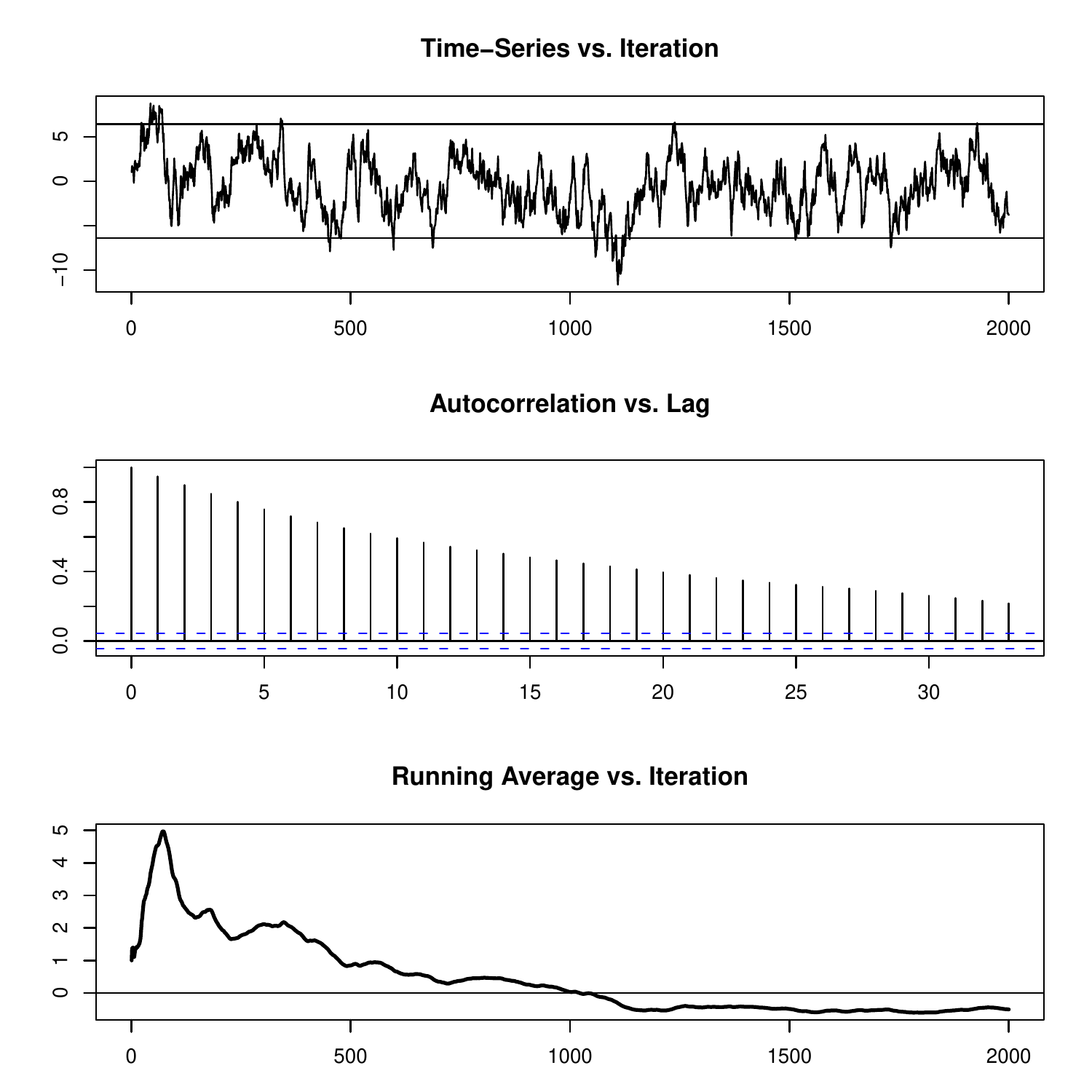}

\FloatBarrier

\subsection{Running MCSEs for Expectations}

The following chunk of code creates the plot of the running MCSEs and running estimates for the expectations with confidence bounds.  Again, the plot is contained in the document.

\begin{Schunk}
\begin{Sinput}
> rho = rho1
> chain <- chain1
> mean <- mean1
> u.obm <- mean + crit.obm * obm.est1
> l.obm <- mean - crit.obm * obm.est1
> u.bm <- mean + crit.bm * bm.est1
> l.bm <- mean - crit.bm * bm.est1
> ts.plot(mean, main = "Running Average", xlab = "Iteration", ylab = "", 
+     ylim = c(min(l.obm[10:n]), max(u.obm[10:n])), lwd = 2, xlim = c(0, 
+         n))
> abline(h = 0)
> points(iter, u.obm, type = "l", lty = 4, lwd = 2)
> points(iter, l.obm, type = "l", lty = 4, lwd = 2)
\end{Sinput}
\end{Schunk}
\includegraphics{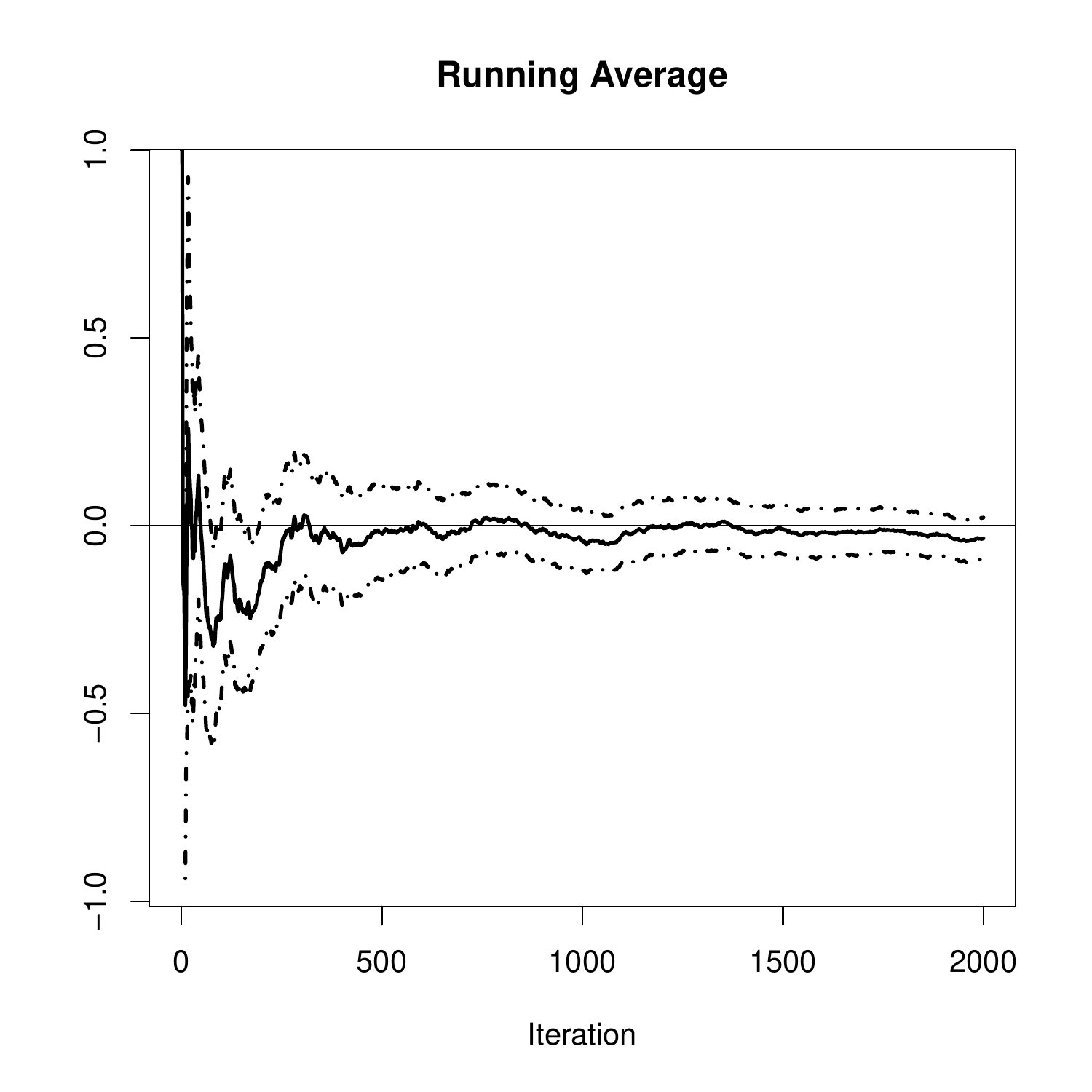}

\begin{Schunk}
\begin{Sinput}
> rho = rho2
> chain <- chain2
> mean <- mean2
> u.obm <- mean + crit.obm * obm.est2
> l.obm <- mean - crit.obm * obm.est2
> u.bm <- mean + crit.bm * bm.est2
> l.bm <- mean - crit.bm * bm.est2
> ts.plot(mean, main = "Running Average", xlab = "Iteration", ylab = "", 
+     ylim = c(min(l.obm[10:n]), max(u.obm[10:n])), lwd = 2, xlim = c(0, 
+         n))
> abline(h = 0)
> points(iter, u.obm, type = "l", lty = 4, lwd = 2)
> points(iter, l.obm, type = "l", lty = 4, lwd = 2)
\end{Sinput}
\end{Schunk}
\includegraphics{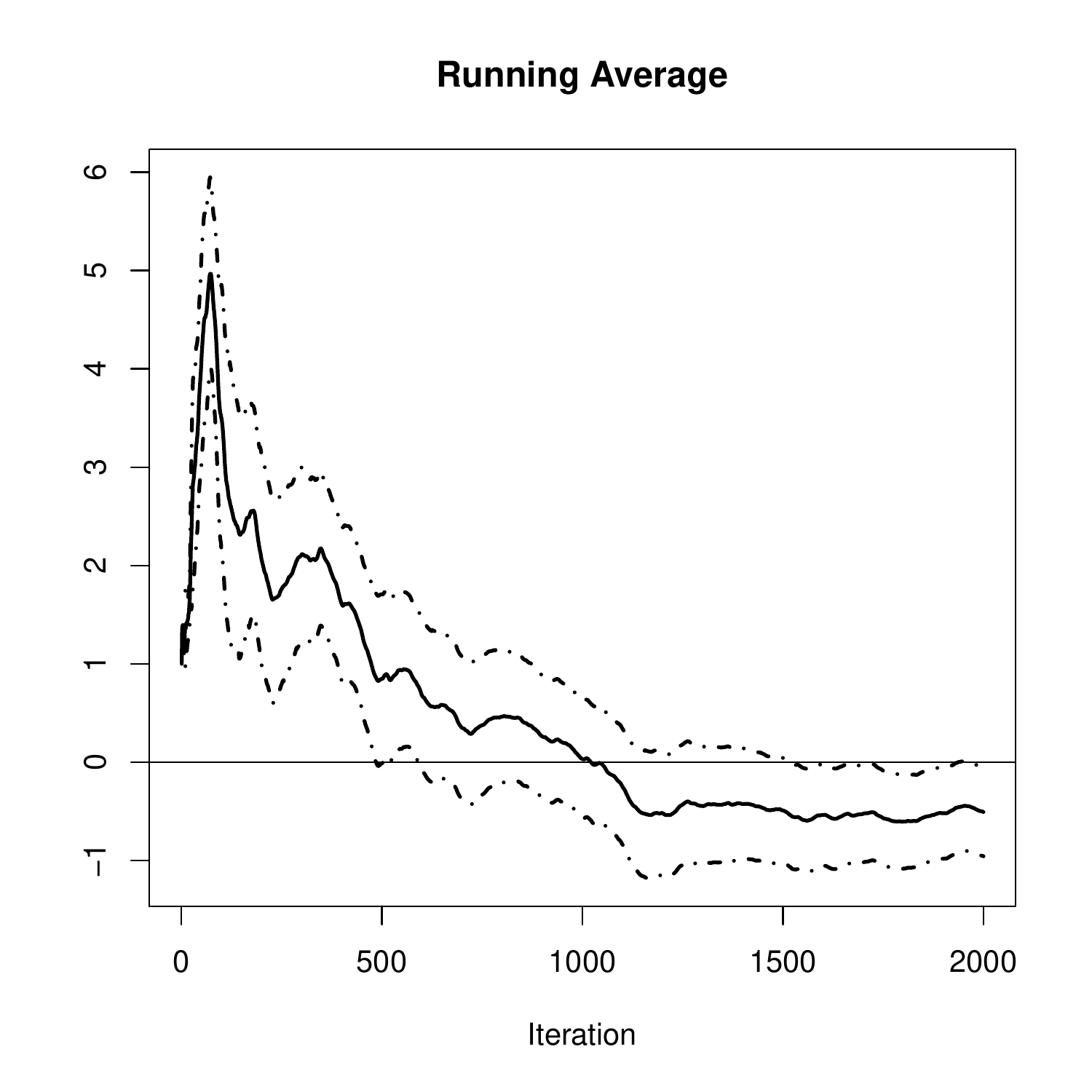}

\FloatBarrier

\subsection{Running Quartile Plots}

Here is the code for the running quartile plots with and without confidence bounds for $\rho \in \{ 0.5 , 0.95 \}$.  There are a total of four chunks of code here, one for each plot.

\begin{Schunk}
\begin{Sinput}
> rho = rho1
> chain <- chain1
> quartiles <- quartile1
> u.sub <- quartiles + rbind(crit.obm, crit.obm) * q.mcse1
> l.sub <- quartiles - rbind(crit.obm, crit.obm) * q.mcse1
> ts.plot(t(quartiles), main = "Running Quartiles", xlab = "Iteration", 
+     ylab = "", lwd = 2, ylim = c(max(u.sub[, 100:n]), min(l.sub[, 
+         100:n])), xlim = c(0, n))
> abline(h = qnorm(0.25, 0, sqrt(1/(1 - rho^2))))
> abline(h = qnorm(0.75, 0, sqrt(1/(1 - rho^2))))
\end{Sinput}
\end{Schunk}
\includegraphics{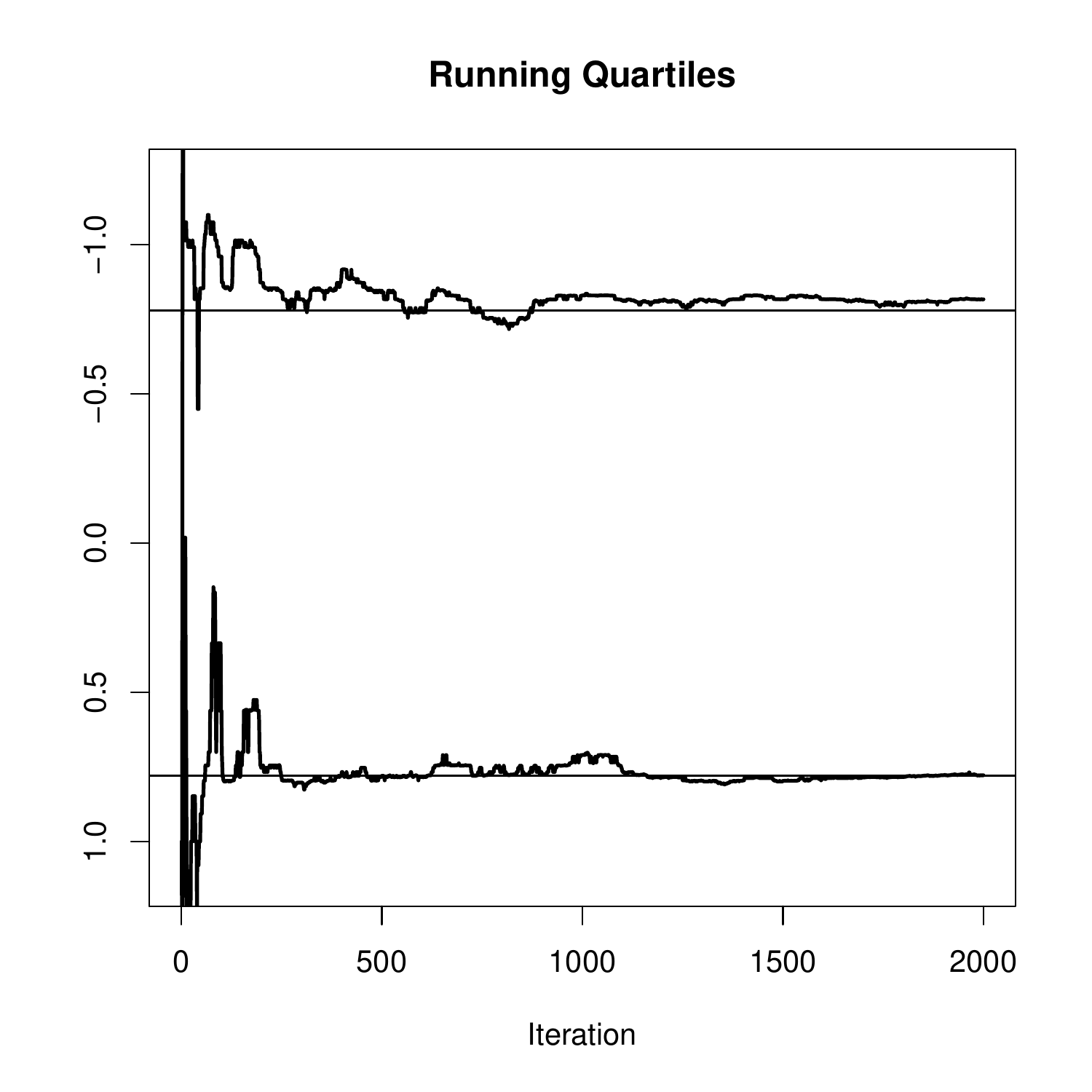}

\FloatBarrier

\begin{Schunk}
\begin{Sinput}
> ts.plot(t(quartiles), main = "Running Quartiles", xlab = "Iteration", 
+     ylab = "", lwd = 2, ylim = c(max(u.sub[, 100:n]), min(l.sub[, 
+         100:n])), xlim = c(0, n))
> points(iter, t(u.sub[1, ]), type = "l", lty = 4, lwd = 2)
> points(iter, t(u.sub[2, ]), type = "l", lty = 4, lwd = 2)
> points(iter, t(l.sub[1, ]), type = "l", lty = 4, lwd = 2)
> points(iter, t(l.sub[2, ]), type = "l", lty = 4, lwd = 2)
> abline(h = qnorm(0.25, 0, sqrt(1/(1 - rho^2))))
> abline(h = qnorm(0.75, 0, sqrt(1/(1 - rho^2))))
\end{Sinput}
\end{Schunk}
\includegraphics{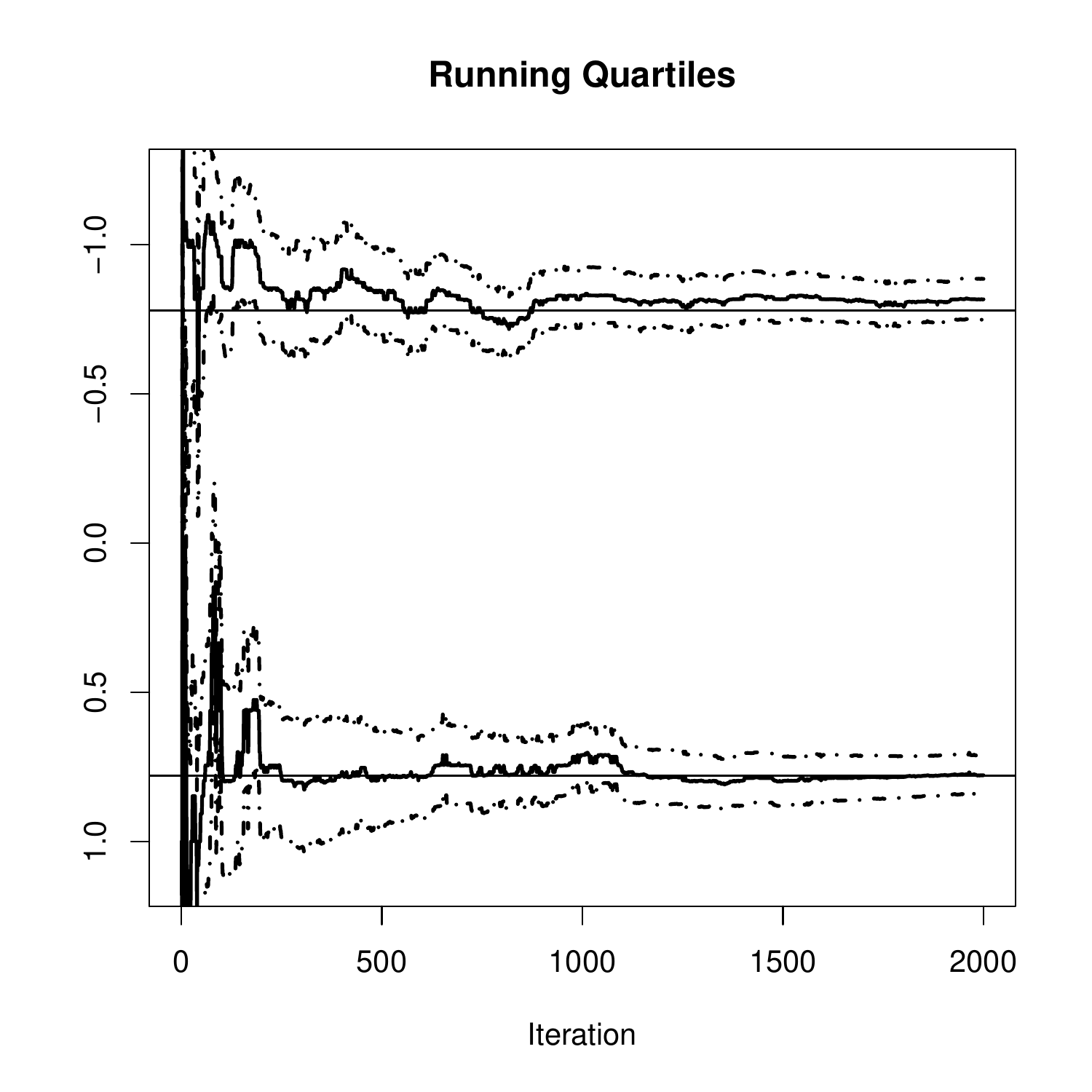}

\FloatBarrier

\begin{Schunk}
\begin{Sinput}
> rho = rho2
> chain <- chain2
> quartiles <- quartile2
> u.sub <- quartiles + rbind(crit.obm, crit.obm) * q.mcse2
> l.sub <- quartiles - rbind(crit.obm, crit.obm) * q.mcse2
> ts.plot(t(quartiles), main = "Running Quartiles", xlab = "Iteration", 
+     ylab = "", lwd = 2, ylim = c(max(u.sub[, 100:n]), min(l.sub[, 
+         100:n])), xlim = c(0, n))
> abline(h = qnorm(0.25, 0, sqrt(1/(1 - rho^2))))
> abline(h = qnorm(0.75, 0, sqrt(1/(1 - rho^2))))
\end{Sinput}
\end{Schunk}
\includegraphics{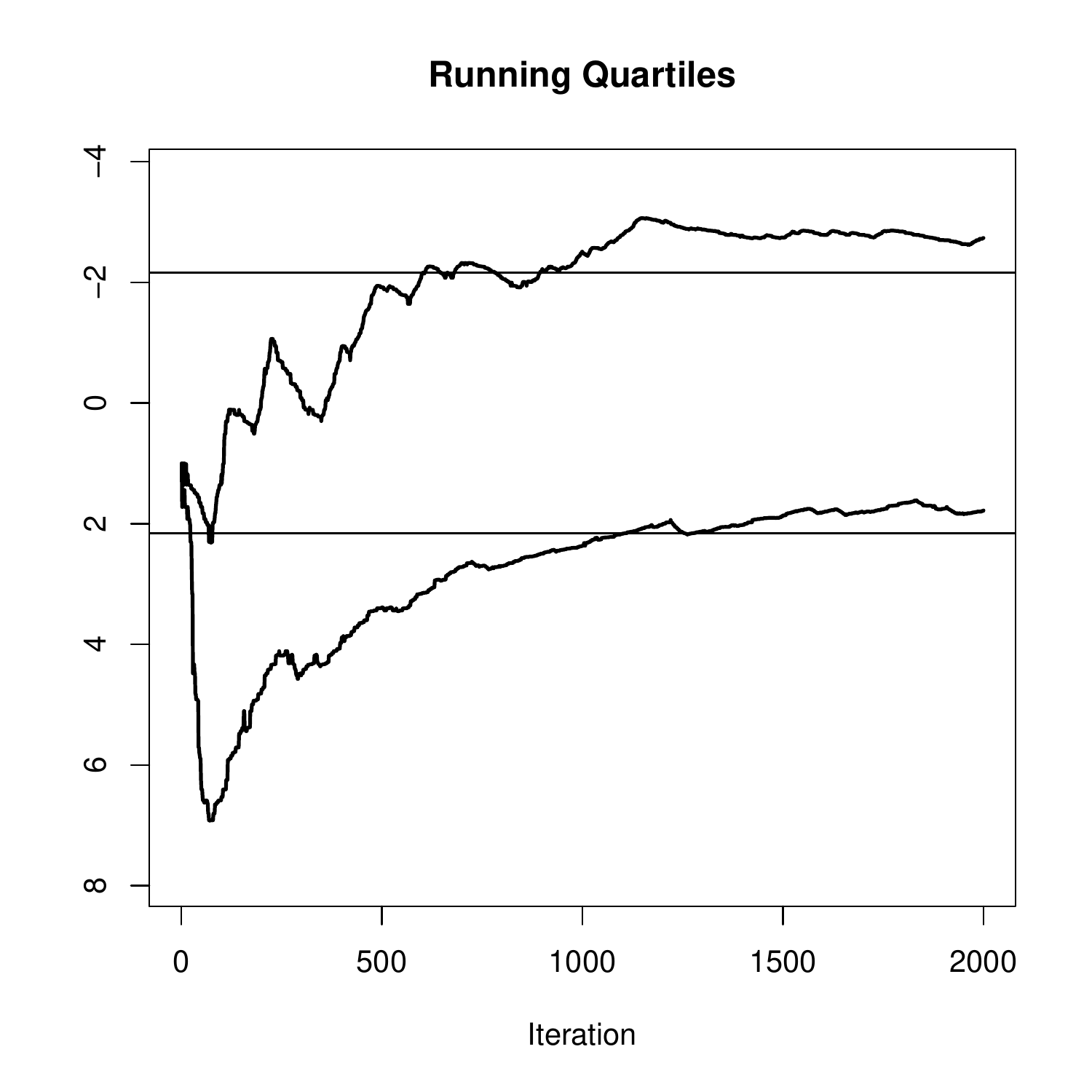}

\FloatBarrier

\begin{Schunk}
\begin{Sinput}
> ts.plot(t(quartiles), main = "Running Quartiles", xlab = "Iteration", 
+     ylab = "", lwd = 2, ylim = c(max(u.sub[, 100:n]), min(l.sub[, 
+         100:n])), xlim = c(0, n))
> points(iter, t(u.sub[1, ]), type = "l", lty = 4, lwd = 2)
> points(iter, t(u.sub[2, ]), type = "l", lty = 4, lwd = 2)
> points(iter, t(l.sub[1, ]), type = "l", lty = 4, lwd = 2)
> points(iter, t(l.sub[2, ]), type = "l", lty = 4, lwd = 2)
> abline(h = qnorm(0.25, 0, sqrt(1/(1 - rho^2))))
> abline(h = qnorm(0.75, 0, sqrt(1/(1 - rho^2))))
\end{Sinput}
\end{Schunk}
\includegraphics{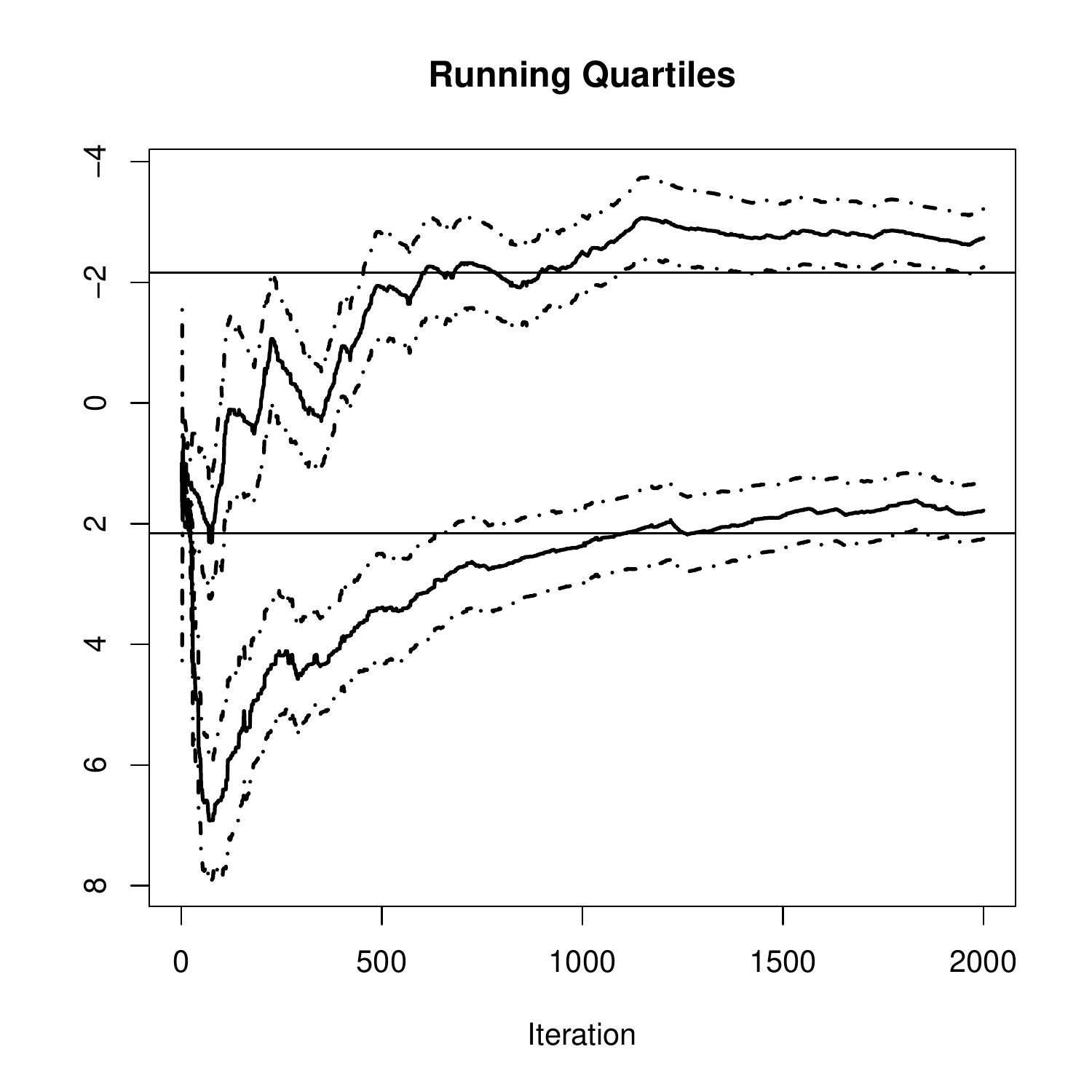}

\FloatBarrier

\subsection{Calculations for Expectations}

The following are calculations reported in the paper based on $n=2000$ iterations in the chain.
\begin{Schunk}
\begin{Sinput}
> signif(mean1[n], 6)
\end{Sinput}
\begin{Soutput}
[1] -0.0336436
\end{Soutput}
\begin{Sinput}
> signif(crit.obm[n] * obm.est1[n], 6)
\end{Sinput}
\begin{Soutput}
[1] 0.0556022
\end{Soutput}
\begin{Sinput}
> signif(mean2[n], 6)
\end{Sinput}
\begin{Soutput}
[1] -0.506755
\end{Soutput}
\begin{Sinput}
> signif(crit.obm[n] * obm.est2[n], 6)
\end{Sinput}
\begin{Soutput}
[1] 0.450564
\end{Soutput}
\begin{Sinput}
> signif(obm.est2[n], 6)
\end{Sinput}
\begin{Soutput}
[1] 0.351458
\end{Soutput}
\end{Schunk}

Here the implementation of fixed width procedures using $\rho = 0.95$.  Here we have added 198000 iterations to the chain for a total of 200000 to spead up the processing time.  If one desired a level of precision smaller than 0.1 it would probably be necessary to add additional iterations to the chain.

\begin{Schunk}
\begin{Sinput}
> chain2 <- ar1.gen(chain2, 198000, rho2, 1)
> chain <- chain2
> half <- crit.obm[n] * obm.est2[n]
> N <- n
> while (half + 1/N > 0.1) {
+     N <- N + 1000
+     b <- floor(sqrt(N))
+     t.OBM <- qt(0.9, (N - b + 1))
+     est.OBM <- mcse(chain[1:N], meth = "OBM")
+     half <- t.OBM * est.OBM
+ }
> N
\end{Sinput}
\begin{Soutput}
[1] 60000
\end{Soutput}
\begin{Sinput}
> half
\end{Sinput}
\begin{Soutput}
[1] 0.0996075
\end{Soutput}
\begin{Sinput}
> mean(chain[1:N])
\end{Sinput}
\begin{Soutput}
[1] -0.04415015
\end{Soutput}
\begin{Sinput}
> signif(mean(chain[1:N]), 6)
\end{Sinput}
\begin{Soutput}
[1] -0.0441502
\end{Soutput}
\end{Schunk}

\subsection{Calculations for Quartiles}

Here are the same calculations for the quartiles.  First the caculations based on $n=2000$ iterations for the quartiles.
\begin{Schunk}
\begin{Sinput}
> signif(quartile1[, n], 6)
\end{Sinput}
\begin{Soutput}
      25%       75% 
-0.816573  0.777758 
\end{Soutput}
\begin{Sinput}
> signif(crit.obm[n] * q.mcse1[, n], 6)
\end{Sinput}
\begin{Soutput}
      25%       75% 
0.0685046 0.0648736 
\end{Soutput}
\begin{Sinput}
> signif(quartile2[, n], 6)
\end{Sinput}
\begin{Soutput}
     25%      75% 
-2.73625  1.77970 
\end{Soutput}
\begin{Sinput}
> signif(crit.obm[n] * q.mcse2[, n], 6)
\end{Sinput}
\begin{Soutput}
     25%      75% 
0.480563 0.466388 
\end{Soutput}
\end{Schunk}

Then the fixed width calculations for quantiles with $\rho = 0.95$.  First for individual CIs, the with a Bonferonni correction.  These calculations are not included in the paper and are commented out since the computational time is approximately an hour.  The code is included for future reference.

\begin{verbatim}
chain <- chain2
half <- max(crit.obm[n]*q.mcse2[,n])
N <- n
while(half + 1 / N > .1){
  N <-  N + 2000
  b <- floor(sqrt(N)) # batch size
  t.sub <- qt(.9, (N - b + 1))
  est.sub <- subsampling(chain[1:N])
  half <- max(t.sub*est.sub)
}
N
half
quant(chain[1:N])
t.sub*est.sub

#####
# Then for Bonferonni Correction
#####

t.sub <- t.sub <- qt(.9875, (N - b + 1))
half <- max(t.sub*est.sub)
while(half + 1 / N > .1){
  N <-  N + 2000
  b <- floor(sqrt(N)) # batch size
  t.sub <- qt(.9875, (N - b + 1))
  est.sub <- subsampling(chain[1:N])
  half <- max(t.sub*est.sub)
}
N
half
quant(chain[1:N])
t.sub*est.sub
\end{verbatim}

\section{T-Distribution Example}
Suppose our goal is to estimate the first two moments of a Students $t$ distribution with 4 degrees of freedom and having density
\[
m (x) = \frac{3}{8} \left( 1 + \frac{x^{2}}{4}\right)^{-5/2} 
\]
Obviously, there is nothing about this that requires MCMC since we can easily calculate that $E_{m} X = 0$ and $E_{m} X^{2} =2 $.  Nevertheless, we will use a data augmentation algorithm based on the joint density
\[
\pi(x, y) = \frac{4}{\sqrt{2\pi}} y^{\frac{3}{2}}  e^{-y \left( 2 + x^{2} /2\right)}
\]
so that the full conditionals are $X | Y \sim \text{N}(0, y^{-1})$ and $Y|X \sim \text{Gamma}(5/2, 2 + x^{2} / 2)$.  Consider the Gibbs sampler that updates $X$ then $Y$ so that a one-step transition looks like $(x', y') \to (x,y') \to (x,y)$.  

\subsection{Markov Chain Sampler}
The following function samples from $p$ iterations in the Markov chain starting from the value $(1,1)$.

\begin{Schunk}
\begin{Sinput}
> t.gen <- function(p, x = 1, y = 1) {
+     loc <- length(x)
+     junk <- double(p)
+     x <- append(x, junk)
+     y <- append(y, junk)
+     for (i in 1:p) {
+         j <- i + loc - 1
+         x[(j + 1)] <- rnorm(1, 0, sqrt(1/y[j]))
+         y[(j + 1)] <- rgamma(1, 5/2, rate = (2 + x[(j + 1)]^2/2))
+     }
+     return(cbind(x, y))
+ }
\end{Sinput}
\end{Schunk}

\subsection{Calculaitons and Plots}
Using the sampler, we can run the simulation and create the necessary plots.
\begin{Schunk}
\begin{Sinput}
> n <- 2000
> iter <- seq(1, n)
> set.seed(100)
> sample <- t.gen((n - 1))
> x <- sample[, 1]
> y <- sample[, 2]
> RB <- 1/y
> x.mean <- cumsum(x)/seq(along = x)
> x2.mean <- cumsum(x^2)/seq(along = x)
> RB.mean <- cumsum(1/y)/seq(along = y)
\end{Sinput}
\end{Schunk}

\begin{Schunk}
\begin{Sinput}
> ts.plot(x.mean, main = "First Moment", xlab = "Iteration", ylab = "", 
+     lwd = 2, xlim = c(0, n))
> abline(h = 0)
> abline(h = 0, lwd = 2, lty = 4)
> legend("topright", c("Std", "RB"), lty = c(1, 4), lwd = 2)
\end{Sinput}
\end{Schunk}
\includegraphics{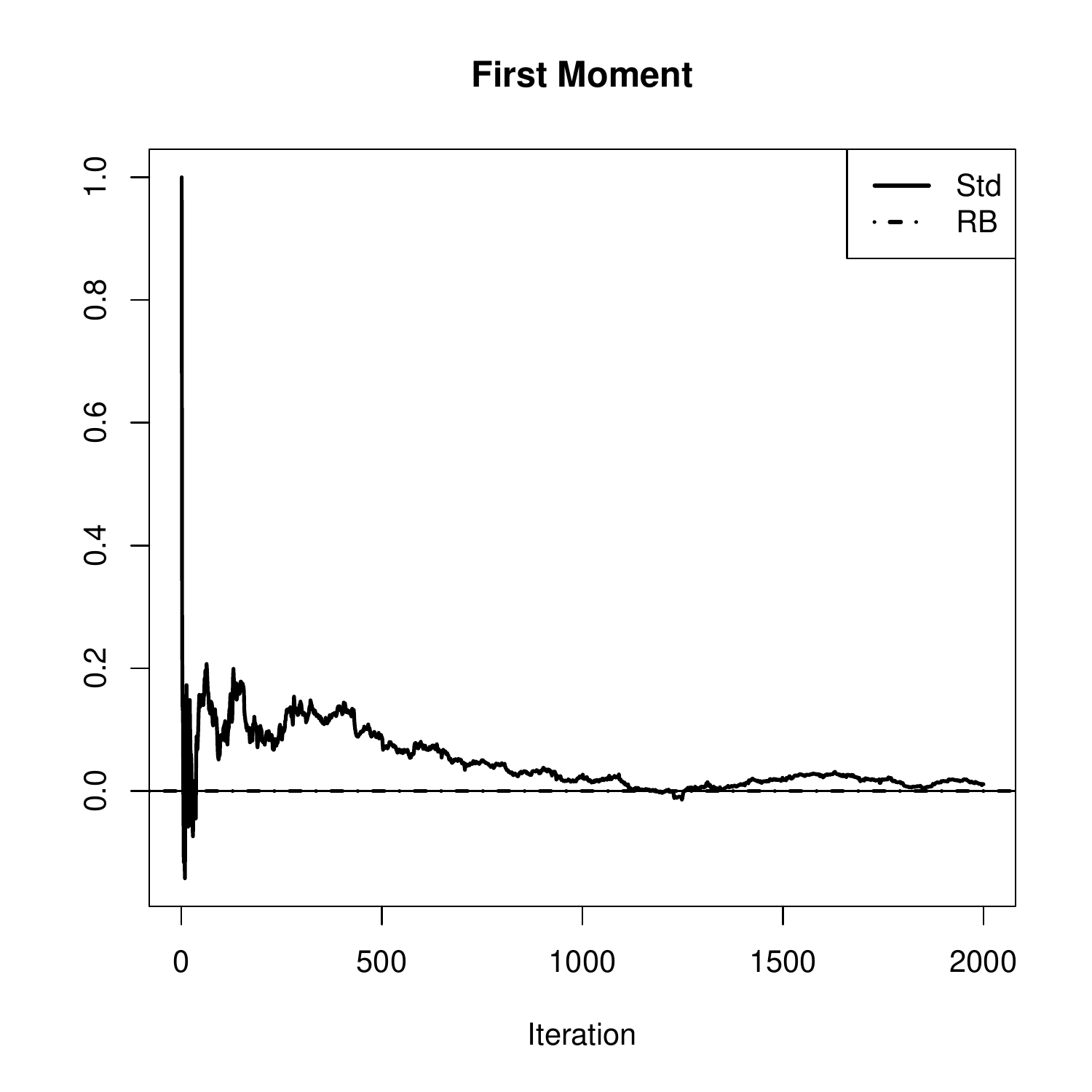}

\begin{Schunk}
\begin{Sinput}
> ts.plot(x2.mean, main = "Second Moment", xlab = "Iteration", 
+     ylab = "", lwd = 2, xlim = c(0, n))
> points(iter, RB.mean, type = "l", lty = 4, lwd = 2)
> abline(h = 2)
> legend("bottomright", c("Std", "RB"), lty = c(1, 4), lwd = 2)
\end{Sinput}
\end{Schunk}
\includegraphics{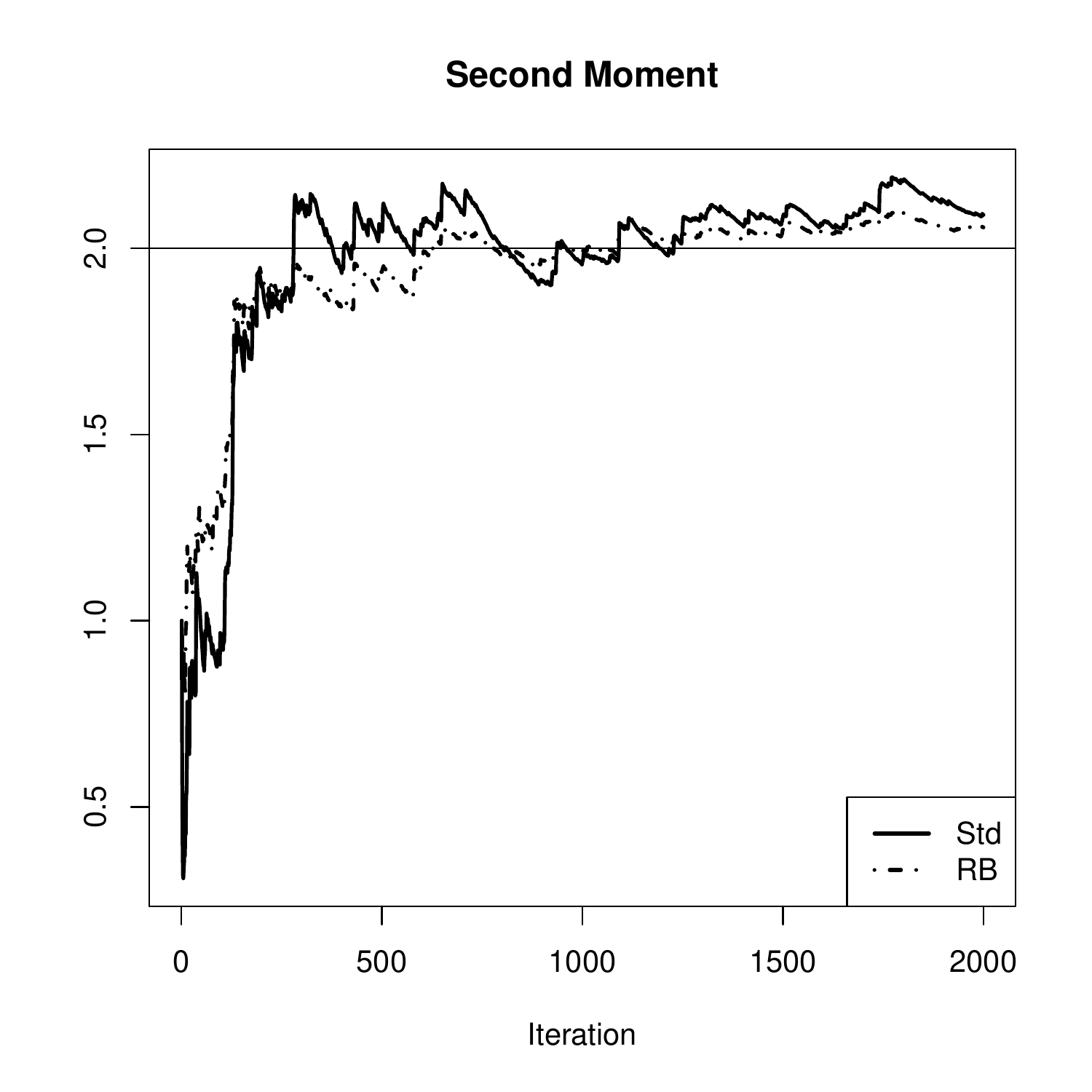}

\FloatBarrier

\subsection{Interval Estimates}
The following chunk of code will calculate interval estimates using the same data from above.
\begin{Schunk}
\begin{Sinput}
> obm.x <- sapply(1:n, function(k) return(mcse(x[1:k], meth = "OBM")))
> obm.RB <- sapply(1:n, function(k) return(mcse(RB[1:k], meth = "OBM")))
> obm.x2 <- sapply(1:n, function(k) return(mcse(x[1:k]^2, meth = "OBM")))
\end{Sinput}
\end{Schunk}

Then we can plot the estimates for the first moment using standard methods.  Notice, there is no uncertainty with the RB estimator in this setting.
\begin{Schunk}
\begin{Sinput}
> upper <- x.mean + crit.obm * obm.x
> lower <- x.mean - crit.obm * obm.x
> ts.plot(x.mean, main = "First Moment", xlab = "Iteration", ylab = "", 
+     lwd = 2, xlim = c(0, n))
> abline(h = 0)
> abline(h = 0, lwd = 2, lty = 4)
> legend("topright", c("Std", "RB"), lty = c(1, 4), lwd = 2)
> points(iter, upper, type = "l", lty = 1, lwd = 1)
> points(iter, lower, type = "l", lty = 1, lwd = 1)
\end{Sinput}
\end{Schunk}
\includegraphics{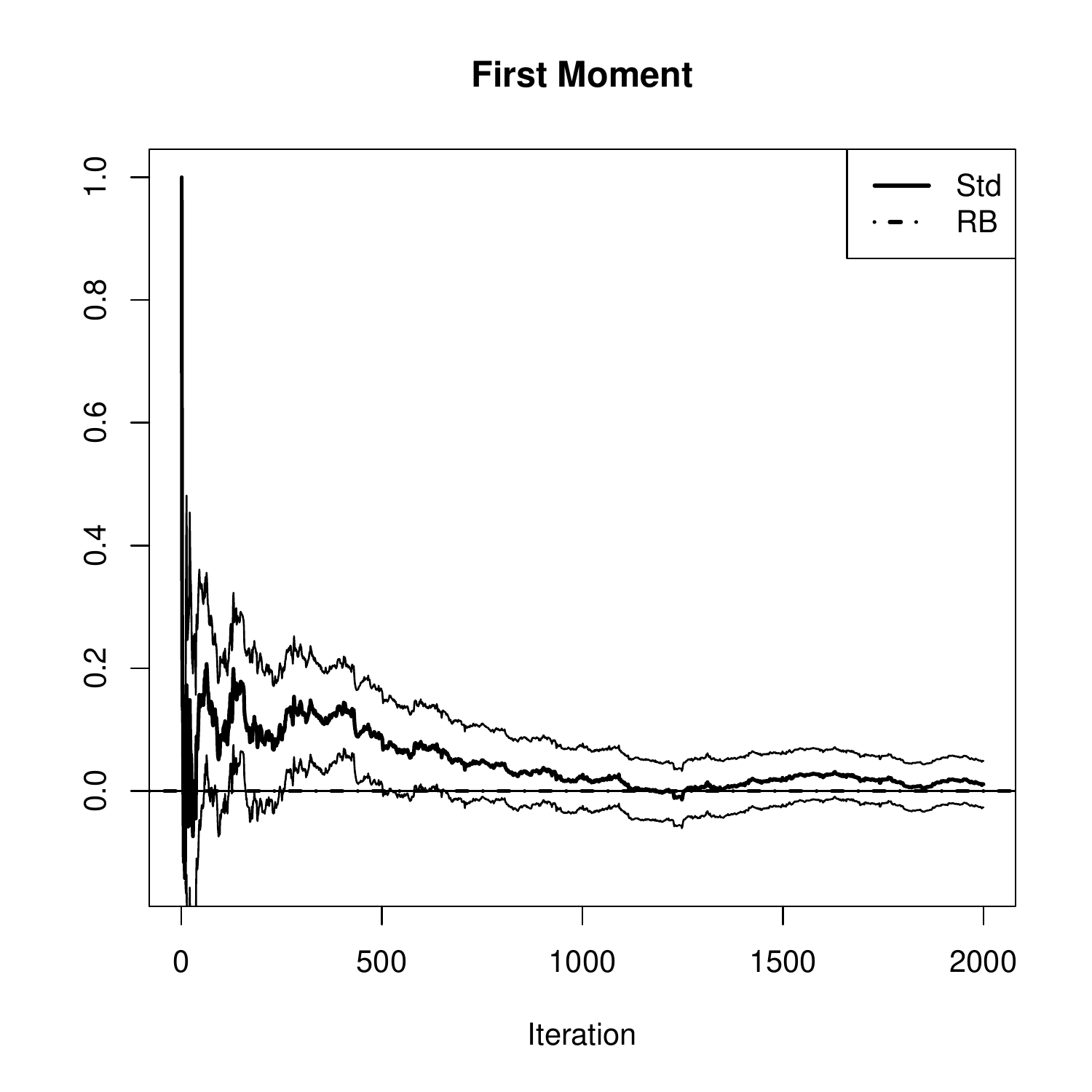}

\FloatBarrier

Now the plot for second moment for both the usual and RB estimator.
\begin{Schunk}
\begin{Sinput}
> upper <- RB.mean + crit.obm * obm.RB
> lower <- RB.mean - crit.obm * obm.RB
> upper.x2 <- x2.mean + crit.obm * obm.x2
> lower.x2 <- x2.mean - crit.obm * obm.x2
> ts.plot(x2.mean, main = "Second Moment", xlab = "Iteration", 
+     ylab = "", lwd = 2, xlim = c(0, n), ylim = c(min(lower.x2[10:n]), 
+         max(upper.x2[10:n])))
> points(iter, RB.mean, type = "l", lty = 4, lwd = 2)
> abline(h = 2)
> legend("bottomright", c("Std", "RB"), lty = c(1, 4), lwd = 2)
> points(iter, upper, type = "l", lty = 4, lwd = 1)
> points(iter, lower, type = "l", lty = 4, lwd = 1)
> points(iter, upper.x2, type = "l", lty = 1, lwd = 1)
> points(iter, lower.x2, type = "l", lty = 1, lwd = 1)
\end{Sinput}
\end{Schunk}
\includegraphics{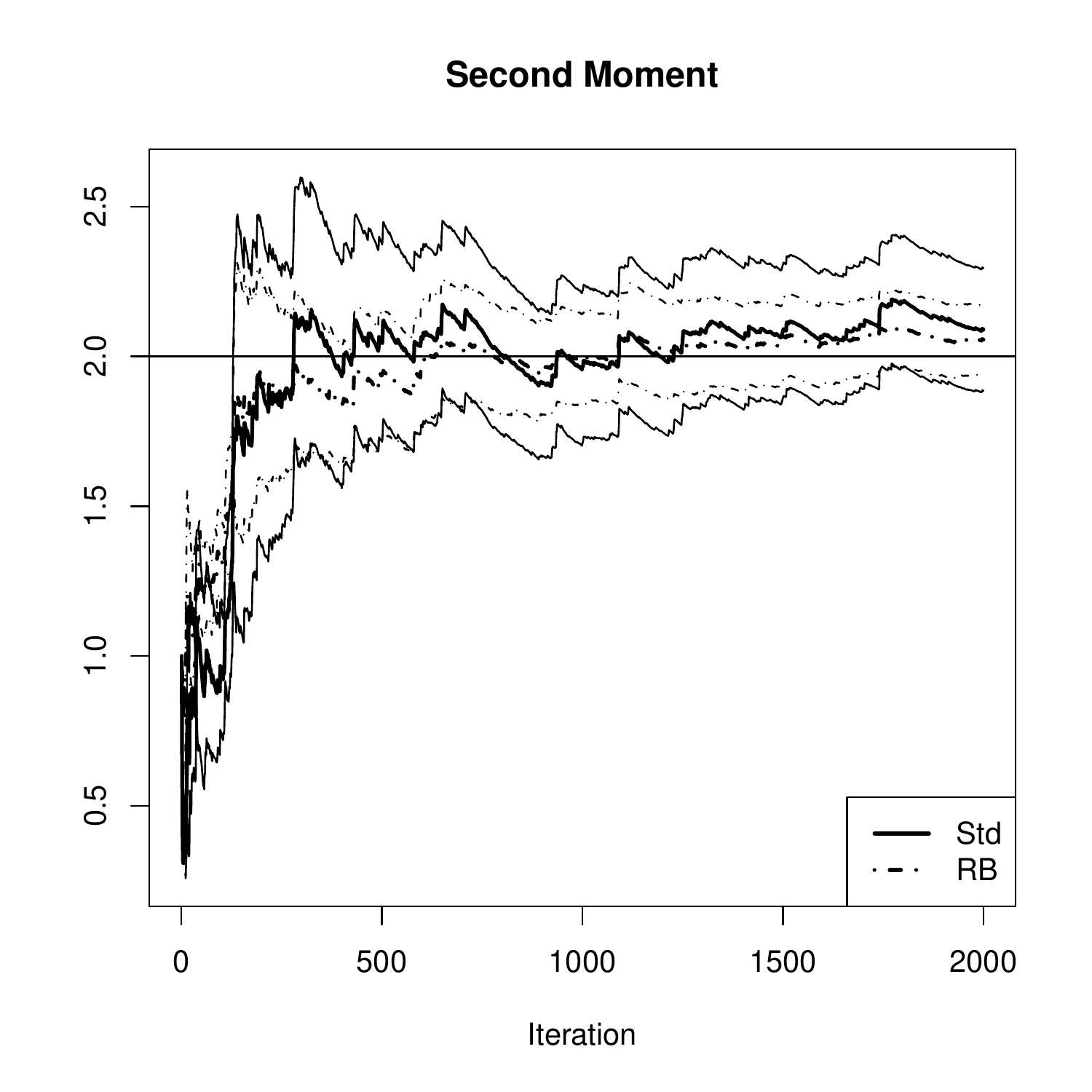}

\section{Estimating Marginals Example}
Suppose $Y_{i} | \mu, \theta \sim \text{N} (\mu, \theta)$ independently for $i=1,\ldots, m$ where $m \ge 3$ and prior $\nu(\mu, \theta) \propto \theta^{-1/2}$.  The target is the posterior density 
\[
\pi(\mu, \theta | y) \propto \theta^{-(m+1)/ 2} e^{- \frac{m}{2\theta} (s^{2} + (\bar{y} - \mu)^{2})}
\]
where $s^{2}$ is the usual biased sample variance.  It is easy to see that $\mu | \theta, y \sim \text{N}(\bar{y}, \theta/m)$ and that $\theta | \mu, y \sim \text{IG} ((m-1)/2, m[s^{2} + (\bar{y} - \mu)^{2}]/2)$.  We will conisder the Gibbs sampler that updates $\mu$ then $\theta$ so that a one-step transition is given by $(\mu', \theta') \to (\mu, \theta') \to (\mu, \theta)$.  We will use this Gibbs sampler to estimate the marginal densities of $\mu$ and $\theta$.

\subsection{Markov Chain Sampler}
The first function provides an observation one step ahead using the Gibbs sampler.  The second function results in $p$ observations from the Markov chain.  
\begin{Schunk}
\begin{Sinput}
> ex2.gibbs <- function(m, t, y.bar = 1, s2 = 4, n = 11) {
+     alpha <- (n - 1)/2
+     beta <- n * (s2 + (y.bar - m)^2)/2
+     t <- rgamma(1, alpha, beta)
+     t <- 1/t
+     m <- rnorm(1, y.bar, sqrt(t/n))
+     cbind(m, t)
+ }
> ex2.gen <- function(mc, p = 100, q = 1, y.bar = 1, s2 = 4, n = 11) {
+     if (is.matrix(mc) == TRUE) {
+         loc <- dim(mc)[1]
+     }
+     else {
+         loc <- 1
+         mc <- t(as.matrix(mc))
+     }
+     junk <- matrix(rep(NA, p * 2), ncol = 2)
+     mc <- rbind(mc, junk)
+     for (i in 1:p) {
+         j <- i + loc - 1
+         mc[(j + 1), ] <- ex2.gibbs(mc[j, 1], mc[j, 2])
+     }
+     return(mc)
+ }
\end{Sinput}
\end{Schunk}

\subsection{Calculations and Settings}

Now suppose $m=11$, $\bar{y}=1$ and $s^{2}=4$.  We simulated $2000$ realizations of the Gibbs sampler starting from $(\mu_{1}, \lambda_{1}) = (1,1)$.  The marginal density plots were created using the default settings for the {\tt density} function in R.  The bivariate density plot was created using R functions  {\tt kde2d} and {\tt persp}.  The resulting posterior is simple, so it is no surprise that the Gibbs sampler has been shown to converge in just a few iterations \citep{jone:hobe:2001}.

\begin{Schunk}
\begin{Sinput}
> n <- 2000
> set.seed(100)
> start <- c(1, 1)
> sample <- ex2.gen(start, (n - 1))
> mu <- sample[, 1]
> theta <- sample[, 2]
\end{Sinput}
\end{Schunk}

Now we can plot the marginal densities and corresponding histograms for $\mu$ and $\theta$.
\begin{Schunk}
\begin{Sinput}
> hist(mu, main = "", xlab = "", freq = F)
> junk <- density(mu)
> points(junk$x, junk$y, type = "l")
\end{Sinput}
\end{Schunk}
\includegraphics{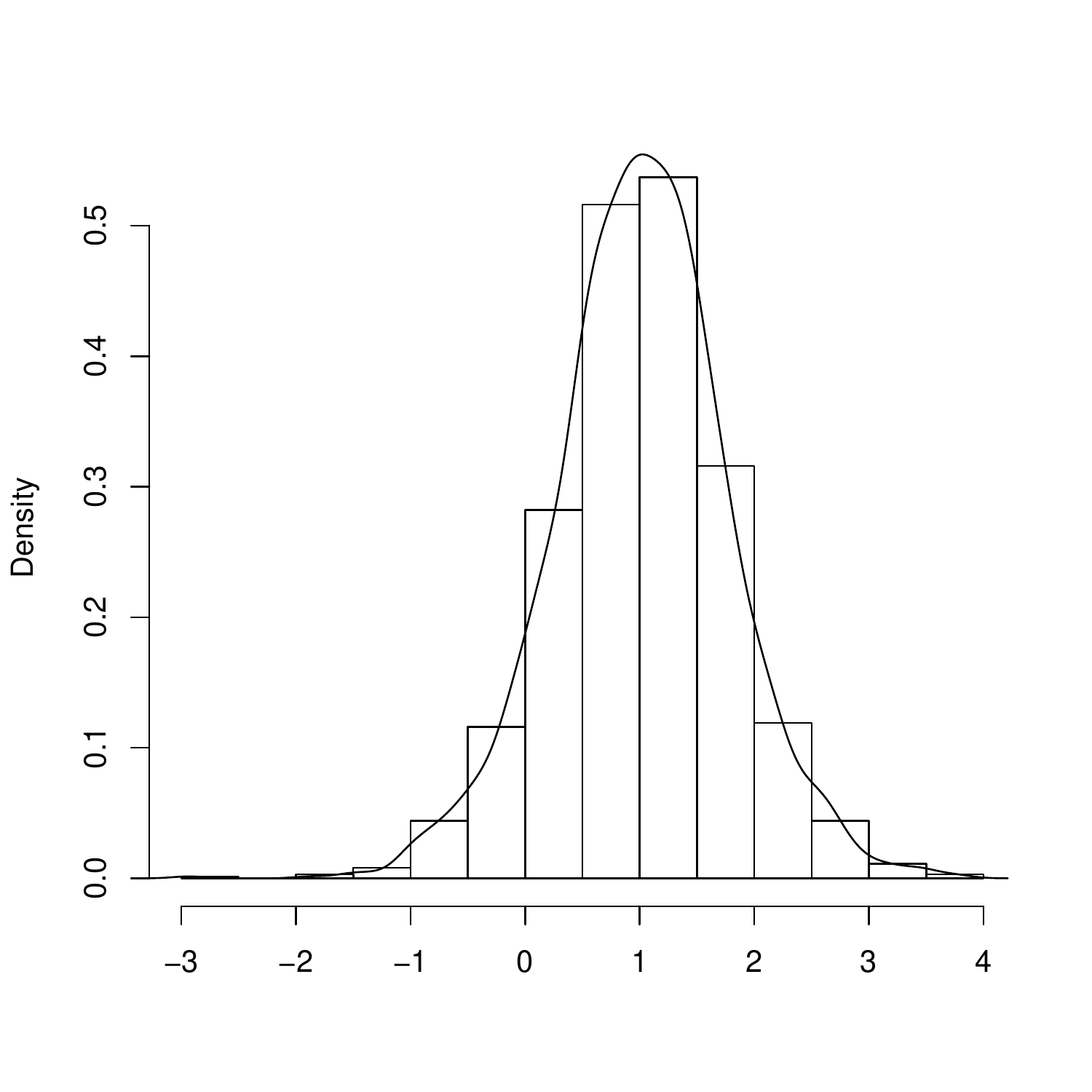}

\begin{Schunk}
\begin{Sinput}
> hist(theta, main = "", xlab = "", freq = F, ylim = c(0, 0.18))
> junk <- density(theta)
> points(junk$x, junk$y, type = "l")
\end{Sinput}
\end{Schunk}
\includegraphics{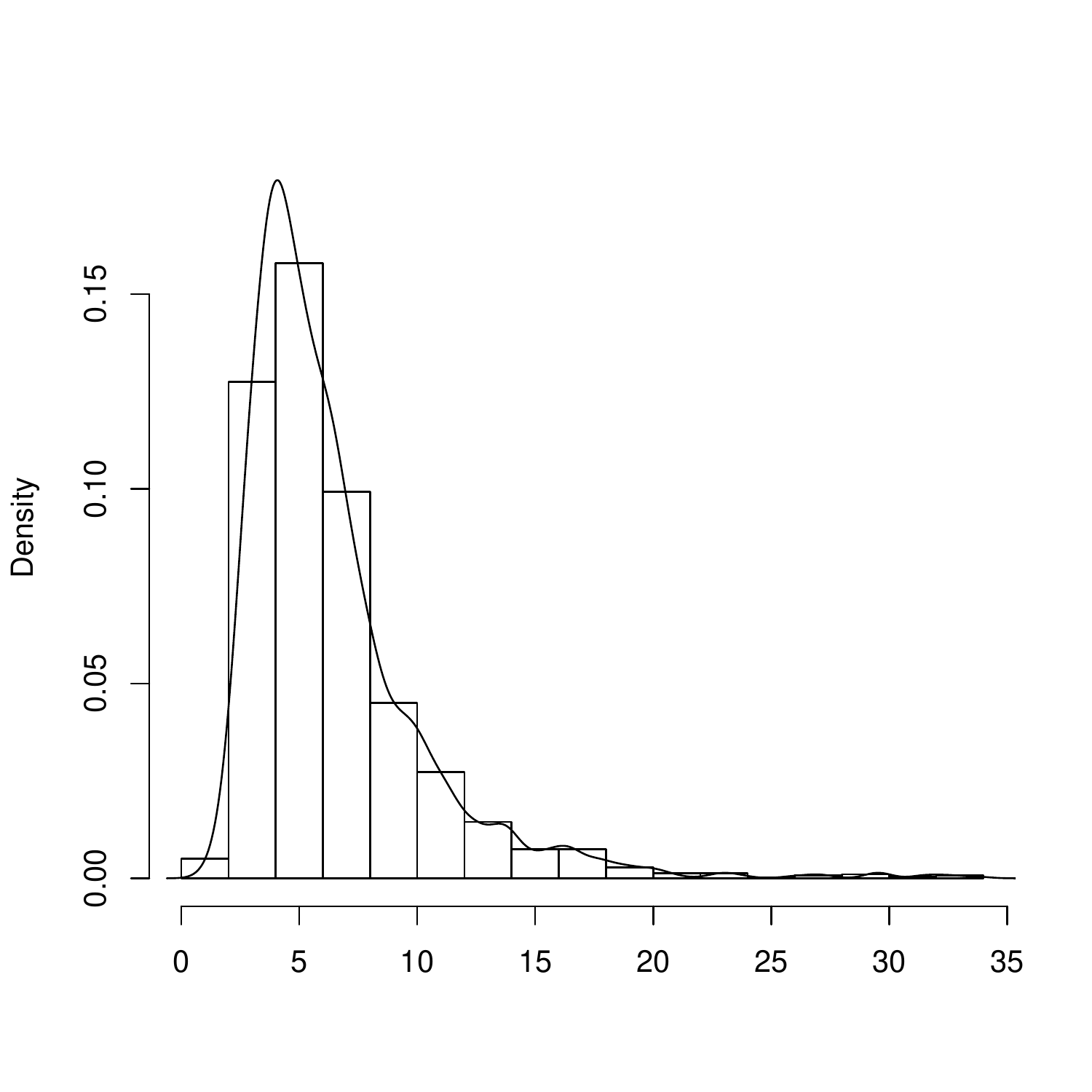}

\FloatBarrier

Now we can plot the bivariate density plot.
\begin{Schunk}
\begin{Sinput}
> library(MASS)
> junk <- kde2d(mu, theta, n = 50, lims = c(-1.5, 3.5, 1, 15))
> persp(junk, theta = 120, phi = 30, xlab = "mu", ylab = "theta", 
+     zlab = "", expand = 0.6)
\end{Sinput}
\end{Schunk}
\includegraphics{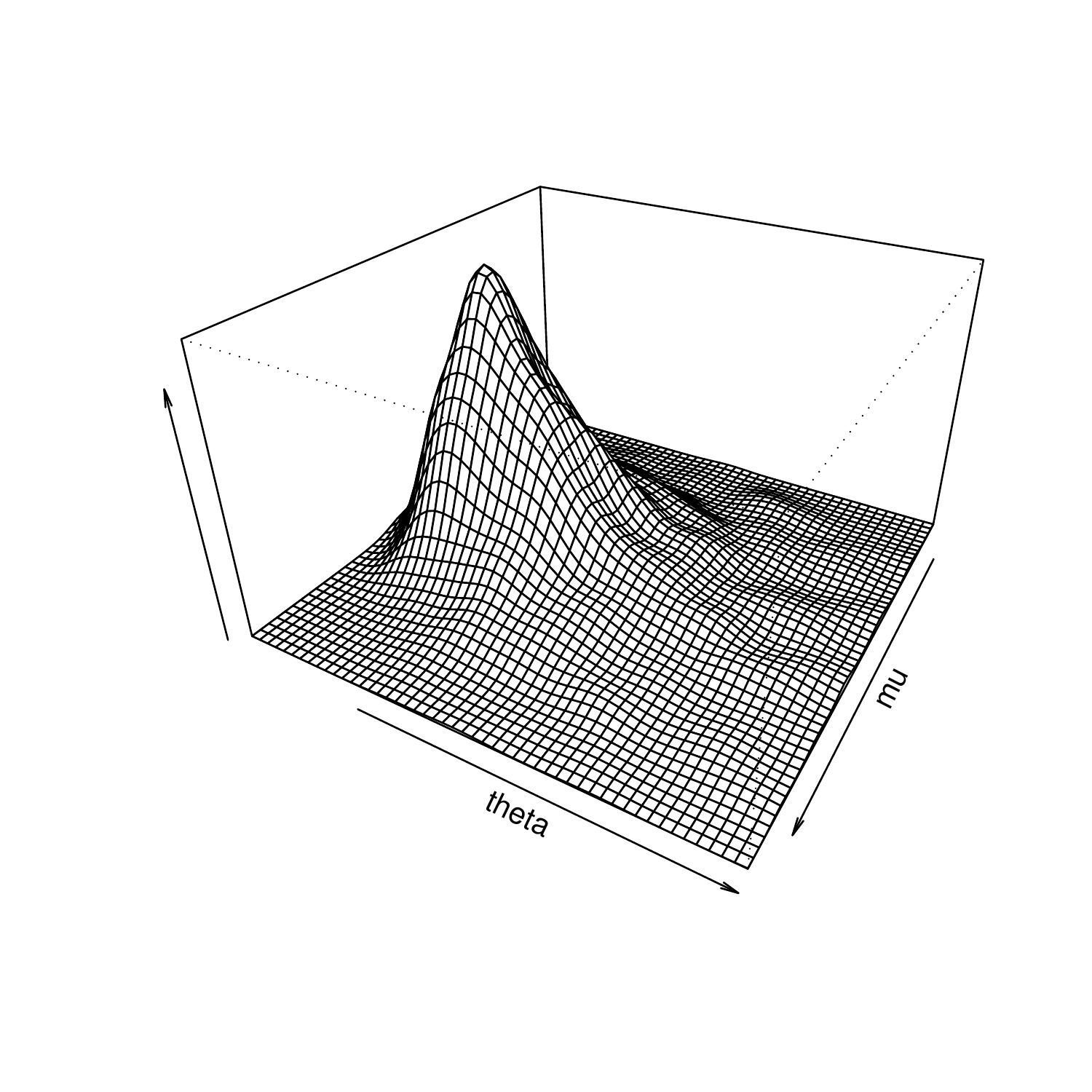}

\FloatBarrier

\subsection{Alternative Curve Estimation}
A clever technique for estimating a marginal is based on the same idea as RB estimators \citep{wei:tann:1990}.  The following chunk of code implements this for the current example for the marginal of $\mu$.
\begin{Schunk}
\begin{Sinput}
> hist(mu, main = "", xlab = "", freq = F)
> junk <- density(mu)
> points(junk$x, junk$y, type = "l")
> evals <- seq(-3, 4, 0.01)
> norm.evals <- dnorm(evals, 1, sqrt(mean(theta)/11))
> points(evals, norm.evals, type = "l", lty = 4, lwd = 2)
\end{Sinput}
\end{Schunk}
\includegraphics{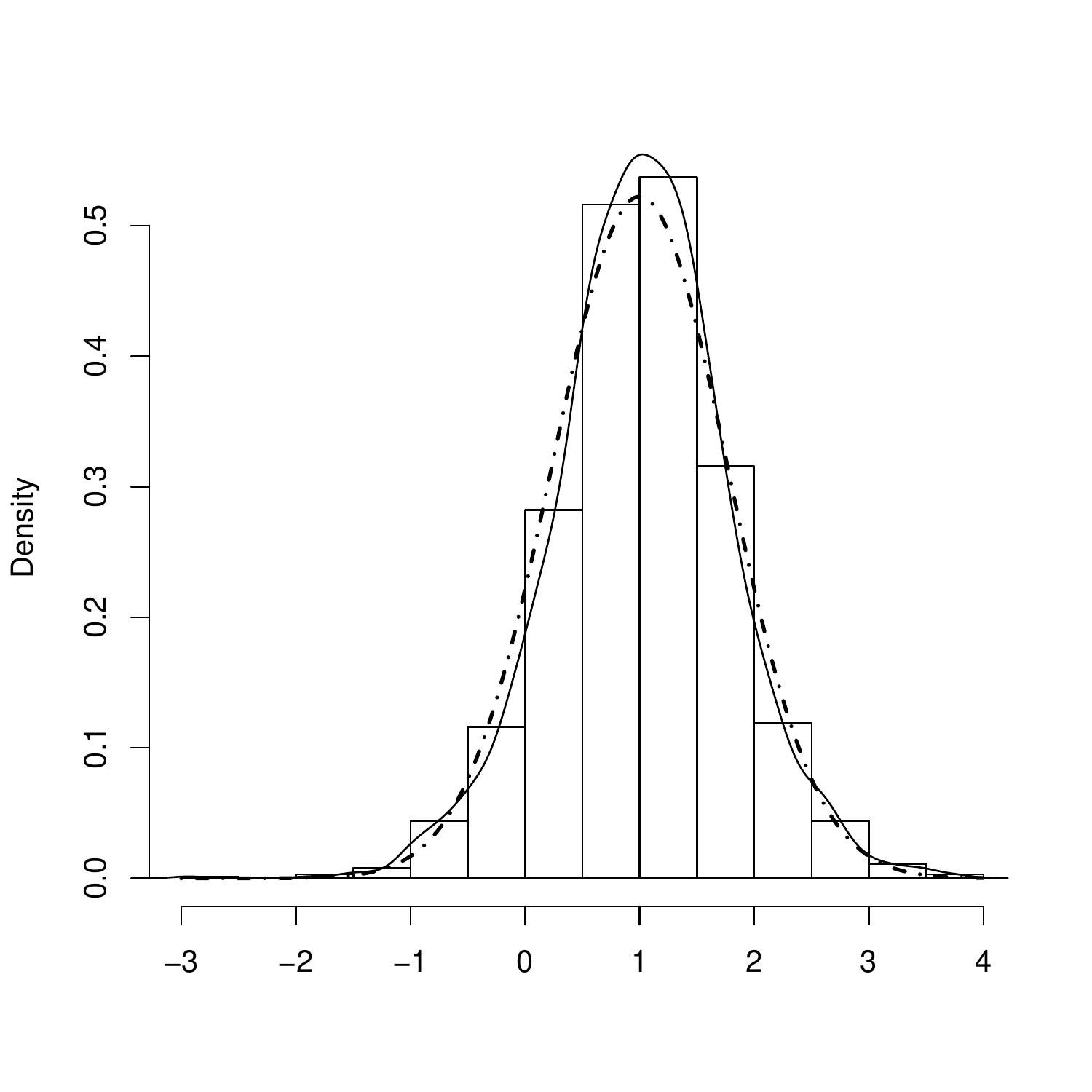}

\bibliographystyle{apalike} 
\bibliography{ref}

\end{document}